\documentclass[fleqn,usenatbib]{mnras}

\usepackage{newtxtext,newtxmath}
\usepackage[T1]{fontenc}

\DeclareRobustCommand{\VAN}[3]{#2}
\let\VANthebibliography\thebibliography
\def\thebibliography{\DeclareRobustCommand{\VAN}[3]{##3}\VANthebibliography}

\usepackage{graphicx}	% Including figure files
\usepackage{amsmath}	% Advanced maths commands
\title[Photometric redshifts for quasars]{Photometric redshifts for quasars from WISE-PS1-STRM}

\author[S. Kunsági-Máté et al.]{
Sándor Kunsági-Máté$^{1}$\thanks{E-mail: skunsagimate@student.elte.hu},
Róbert Beck$^{1}$,
István Szapudi$^{2}$,
István Csabai$^{1}$,
\\
$^{1}$Department of Physics of Complex Systems, ELTE Eötvös Loránd University, Pázmány Péter sétány 1/a, Budapest 1117, Hungary\\
$^{2}$Institute for Astronomy, University of Hawaii, 2680 Woodlawn Drive, Honolulu, HI, 96822\\
}

\date{Accepted XXX. Received YYY; in original form ZZZ}

\pubyear{2022}

\begin{document}
\label{firstpage}
\pagerange{\pageref{firstpage}--\pageref{lastpage}}
\maketitle

\begin{abstract}
Three-dimensional wide-field galaxy surveys are fundamental for cosmological studies. For higher redshifts ($z \gtrsim 1.0$), where galaxies are too faint, quasars still trace the large-scale structure of the Universe. 
Since available telescope time limits spectroscopic surveys, photometric methods are efficient for estimating redshifts for many quasars. Recently, machine learning methods are increasingly successful for quasar photometric redshifts,  however, they hinge on the distribution of the training set. Therefore a rigorous estimation of reliability is critical. 
We extracted optical and infrared photometric data from the cross-matched catalogue of the WISE All-Sky and PS1 3$\pi$ DR2 sky surveys. We trained an XGBoost regressor and an artificial neural network on the relation between color indices and spectroscopic redshift. We approximated the effective training set coverage with the K nearest neighbors algorithm. We estimated reliable photometric redshifts of 2,879,298 quasars which overlap with the training set in feature space. We validated the derived redshifts with an independent, clustering-based redshift estimation technique. The final catalog is publicly available. 

%Three-dimensional wide field surveys are fundamental components of cosmological studies, where quasars are known as good tracers of the large-scale structure of the Universe, especially for higher redshifts ($z \gtrsim 1.0$).
%Due to the substantial telescope time requirements of spectroscopic surveys the estimation of the redshift for a large number of quasars is only possible using photometric measurements. Machine learning based methods are very popular in this field, however they strongly rely on the distribution of the training set. Therefore the reliability of the redshift estimation on the inference set must be validated in a rigorous way.
%We used the cross-matched catalogue of the WISE All-Sky and PS1 3$\pi$ DR2 sky surveys to get optical and infrared photometric data. We trained an XGBoost regressor and an artificial neural network on the relation between the color indices and the spectroscopic redshift. We approximated the effective training set coverage with K nearest neighbors algorithm. We estimated reliable photometric redshifts of 2,879,298 quasars which overlap with the taining set in feature space. The resulting catalog is made publicly available where the derived redshifts have also been validated with a completely independent, clustering-based redshift estimation technique.

\end{abstract}


%%%%%%%%%%%%%%%%%%%%%%%%%%%%%%%%%%%%%%%%%%%%%%%%%%

%%%%%%%%%%%%%%%%% BODY OF PAPER %%%%%%%%%%%%%%%%%%

\section{Introduction}

The three-dimensional distribution of objects of our Universe is a crucial input in several cosmological studies. Although the known optically observable edge of the Universe is about more than 33 billion light years ($z > 11$) away from us (\cite{distantgal}), we still have relatively dense redshift measurements only of a near ($z < 1.$) region which is a tiny part of the whole observable volume. Precise redshift determination needs spectroscopic surveys such as SDSS (\citet{sdss1}). However, faraway objects get so faint that the measurement of their spectra cannot be done or would need extremely long exposure. Due to these difficulties most of the recent and upcoming sky surveys (DES \cite{DES}, LSST \cite{lsst}, WISE \cite{Wright2010}, PanSTARRS \cite{panstarrs}) provide imaging data only. Several methods have been therefore developed in the last decades to create effective models that are able to estimate the redshift from observed fluxes measured in broadband filters. These approaches can be categorized into two groups, namely the template-fitting (\cite{Benitez2000}, \cite{bolzonella}, \cite{Csabai2000}, \cite{ilbert}, \cite{Coe2006}, \cite{Brammer2008}, \cite{leistadt}, \cite{beck1}) and the empirical (\cite{wadadekar}, \cite{Boris2007}, \cite{miles}, \cite{Budavari2009}, \cite{Carliles2010}, \cite{omill}, \cite{krone}, \cite{elliott}, \cite{hogan}) methods. Since the template fitting method relies on a physical model, it generalizes/extrapolates typically better than the empirical methods. However the empirical, mostly Machine Learning methods are better at interpolating within a subregion in the feature space specified by the spectroscopic training sample and so to avoid errors from unknown observational biases. In this work we used two empirical models, XGBoost and Artificial Neural Networks to provide reliable photometric redshift estimations for close to three millions of quasars detected in the PS1-WISE cross-match catalog. As a comparison one of the most recent quasar catalog with photometric redshifts contains about one million of objects (\cite{kidscatalog}). There are other photometric redshift catalogs as well consisting however much less quasars such as \cite{Yang} and \cite{wu}, where the relevance of the infrared bands in the redshift estimation have been confirmed.\\
This paper is organized as follows: in Section \ref{data} we give all the necessary details of the used data, in Section \ref{methods} we give all the information about the used methods, in Section \ref{results} we present and discuss the results and finally in Section \ref{conclusions} we summarize our work.

\section{Data}\label{data}

We used the cross-matched catalogue of the WISE All-Sky and PS1 3$\pi$ DR2 sky surveys presented by Beck et al. 2022 (submitted). They provided a highly accurate source classification with 97.67\% purity and 94.28\% completeness with respect to quasars. After we successfully trained our photo-z model on the spectroscopically identified quasars we applied the model on the quasar candidates found by Beck et al. 2022 (submitted). The Pan-STARRS survey performed broad-band photometric measurements of about three quarters of the sky mainly in the optical regime using the g, r, i, z, y filters (\citet{tonry}, \citet{panstarrs}, \citet{Magnier2020}, \citet{Magnier2020b}, \citet{Magnier2020c}, \citet{Waters2020}). We used the Kron and PSF (Point-spread function) magnitudes of the objects measured in the mentioned filters. The WISE survey scanned the full sky in four infrared photometric bands (W1, W2, W3, W4) having effective wavelengths of 3.4, 4.6, 12 and 22 $\mu$m, respectively (\citet{Wright2010}). Regarding the high noise level and the relatively large number of missing error estimates of the W3, W4 filters, we only used the measurements obtained in the W1 and W2 filters. We selected the profile fitting photometry -- which essentially fits a point-spread function on the data -- as well as the aperture magnitudes related to 8.25'' radius circular apertures. Finally, we determined the \textit{color indices}, namely the magnitude differences of the neighboring filters by pairing the PS1 Kron and WISE aperture magnitudes as well as the PS1 PSF and WISE PSF magnitudes to each other. Hence, we ended up with a 12 dimensional feature space. Note that due to model extrapolation considerations it is very important to rely on such input parameters (in our case color indices) that are less sensitive to the actual magnitudes since the spectroscopic training set is typically brighter than the photometric inference set. The PS1 magnitudes have been corrected for the galactic dust extinction using the related extinction coefficients ($\alpha_g = 3.172$,$\alpha_r = 2.271$,$\alpha_i = 1.682$,$\alpha_z = 1.322$,$\alpha_y = 1.087$) and the E(B-V) dust extinction values of a map that is based on PS1 observations of Milky Way stars (\citet{Schlafly2014}). For spectroscopic redshifts we used SDSS data (\citet{York2000}, \citet{Lyke2020}), where the derived training set consisted of 346,691 quasars.

\section{Methods}\label{methods}

\begin{figure}
	\includegraphics[width=\columnwidth]{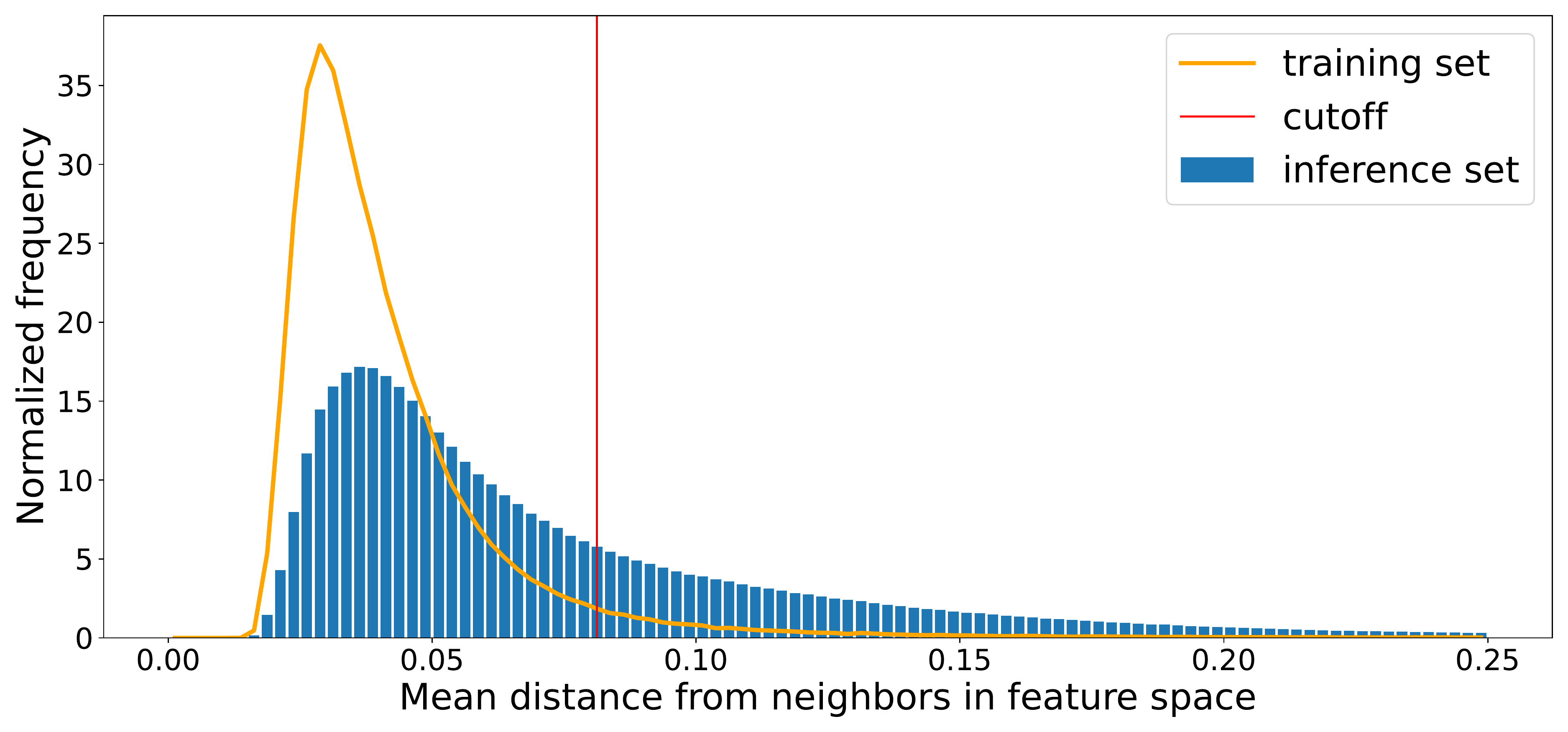}
    \caption{Frequency distribution of the mean distance of each data point from their neighbors in the 12-dimensional normalized feature space with respect to the spectroscopic data set (orange continuous line). Blue bars denote the distribution of the average distance measured between the inference data points and their spectroscopic neighbors. The red vertical line indicates the cut off distance value which corresponds to the 95th percentile of the spectroscopic data set.}
    \label{fig:knn_dist}
\end{figure}

First of all we estimated the training set coverage to provide a well defined boundary in the normalized feature space\footnote{We transformed all of the features to a distribution having a zero mean and a standard deviation of 1.} where our model predictions are reliable. To do this we searched for the 20 nearest neighbors in the 12-dimensional feature space using the ball tree algorithm (\citet{balltree}). We then calculated the mean distance from the neighbors for each data point and investigated its distribution (see Figure \ref{fig:knn_dist}).

The red vertical line indicates the cut off distance value which corresponds to the 95th percentile. Next, using the previously determined K nearest neighbors model we calculated the distance of each inference data point from the 20 nearest neighbors lying in the training set, and again we calculated the mean of these distances. This way we can accurately determine the overlapping region of the training and inference sets in the feature space. We plotted the distribution of the distances denoted by blue bars in Figure \ref{fig:knn_dist}. Altogether 2,879,298 from the total 4,849,611 quasar candidates in the inference set are closer to the training set than the cut off value. This means that the photometric redshift estimation based on our training set is the most reliable on this subset of the quasar candidates, otherwise we extrapolate into a less represented region. 

We trained an XGBoost regressor (XGB) and an artificial neural network (ANN) on the complex relation between the feature space and the spectroscopic redshifts. We used XGB (\citet{xgb}) as a baseline photo-z model and to measure the feature importances. XGB is a boosting algorithm that was developed on the basis of Gradient Boosting Decision Tree (\citet{gbdt}). The main difference between the two models is that during the training phase XGB uses both of the first and second derivatives of the loss function. We set the number of estimators to 12 and the maximum depth to 15. The final photo-z catalogue was created however using the ANN, that outperformed XGB on the training set. We used four hidden layers each having 512 neurons and Exponential Linear Unit (ELU) activation function. ELU is more advanced than the widely used Rectified Linear Unit (RELU) since it solves the problem of vanishing gradient, while providing lower training times and higher accuracy. To avoid overfitting we added dropout layers after each hidden layer with  a dropout rate of 0.2.

\section{Results}\label{results}

\subsection{Application of XGBoost}

We split the spectroscopic data set into train, test and validation sets using 70\% - 15 \% - 15\% ratios, respectively. To quantify the goodness of the model predictions we used the following metrics ($N$: number of data points, $z_{phot}$: photometric redshift, $z_{spec}$: spectroscopic redshift):

\begin{itemize}
    \item Mean Squared Error (MSE):
    \begin{equation}
        MSE = \frac{1}{N} \sum\limits_{i=0}^N (z_{phot,i} - z_{spec,i})^2
        \label{eq:mse}
    \end{equation}
    \item $\delta z_{norm,i}$:
    \begin{equation}
         \delta z_{norm,i} = \frac{z_{phot,i} - z_{spec,i}}{1 + z_{spec,i}}
        \label{eq:deltaznorm}
    \end{equation}
    \item Median Absolute Deviation of $\delta z_{norm,i}$ (MAD)
    \item Mean Absolute Difference (MeanAD):
    \begin{equation}
        \frac{1}{N} \sum\limits_{i=0}^N |z_{phot,i} - z_{spec,i}|
        \label{eq:meanad}
    \end{equation}
    \item Bias (B):
    \begin{equation}
        \frac{1}{N} \sum\limits_{i=0}^N \delta z_{norm,i}
        \label{eq:bias}
    \end{equation}
    \item Outlier rate (O): fraction of objects, where $|z_{phot,i} - z_{spec,i}| > 3 \sqrt{MSE}$ 
\end{itemize}

\begin{figure}
	\includegraphics[width=\columnwidth]{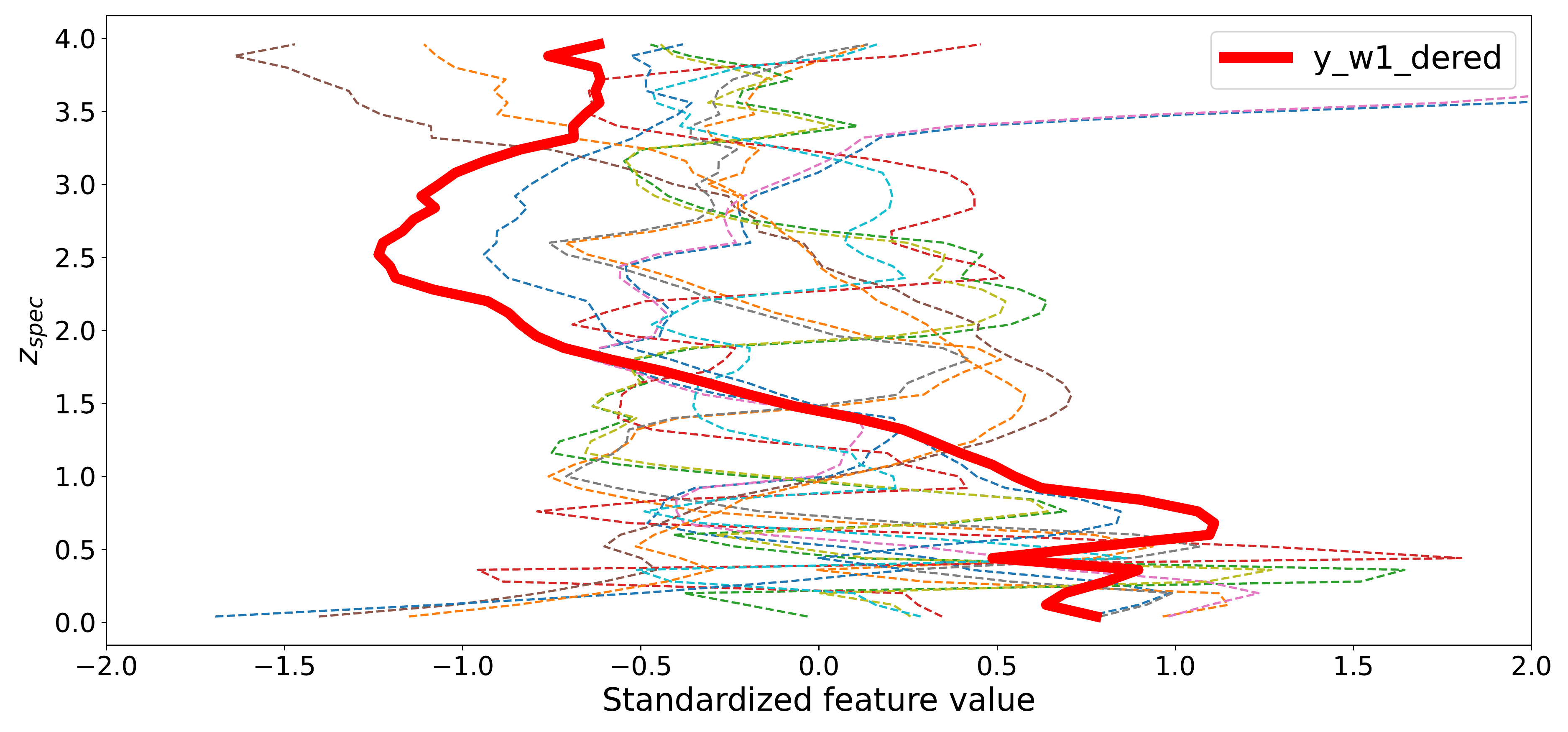}
    \caption{The used redshift bin centers as a function of median feature values. The \texttt{y\_w1\_dered} feature is marked with the red line.}
    \label{fig:colind_z}
\end{figure}

\begin{figure}
	\includegraphics[width=\columnwidth]{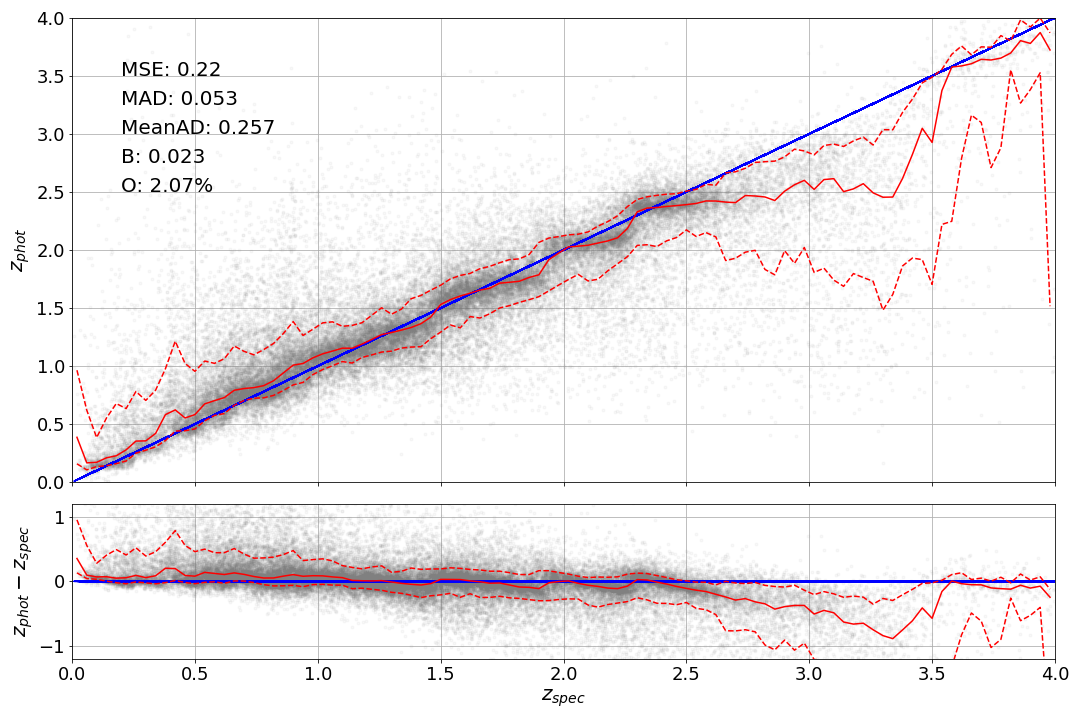}
    \caption{Photometric redshift estimates of XGBoost regressor and the residuals. The red continuous and dashed lines refer to the median and the 68$\%$ confidence interval, respectively.}
    \label{fig:photoz_XGB}
\end{figure}

\begin{figure}
	\includegraphics[width=\columnwidth]{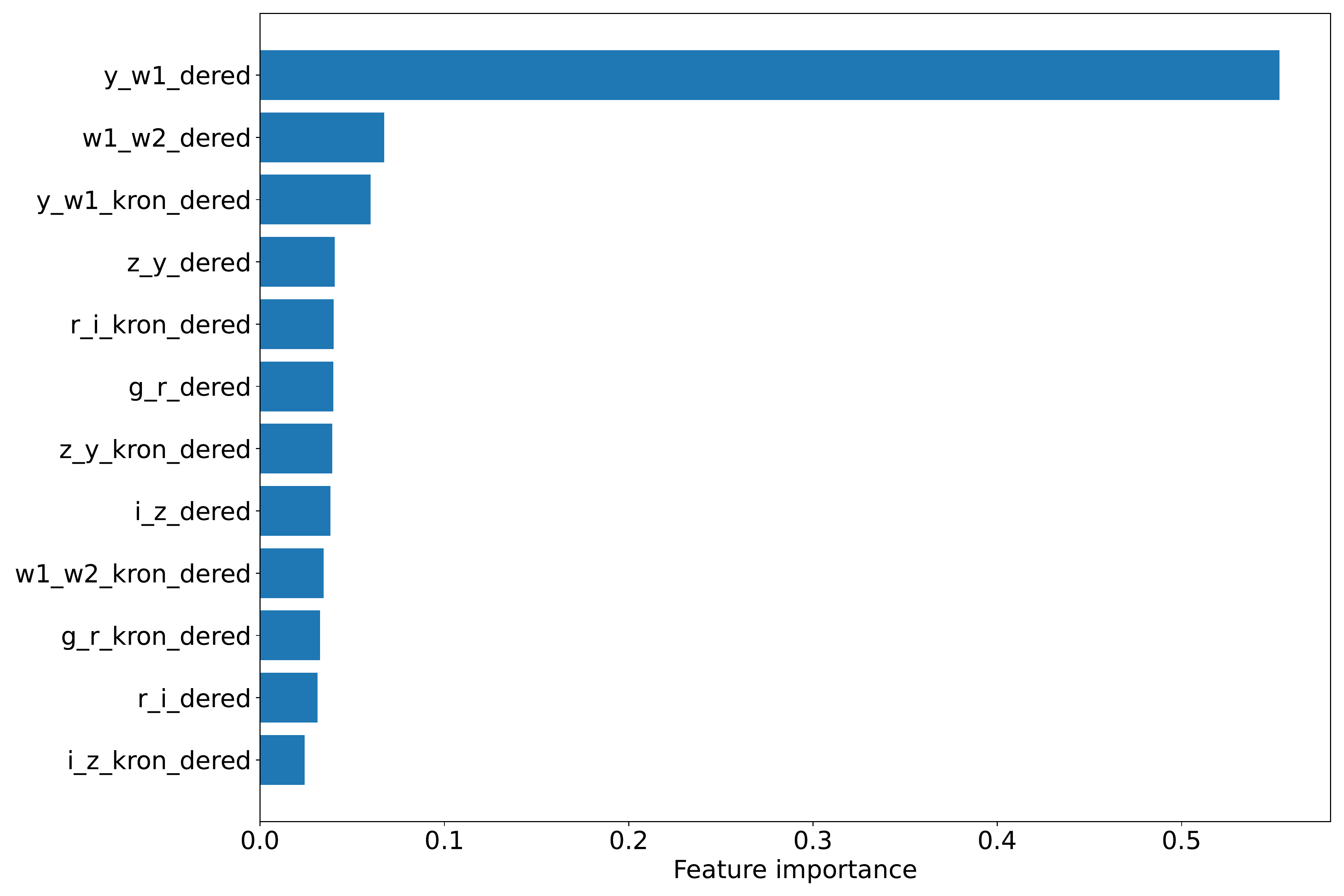}
    \caption{Feature importance value of the different color indices.}
    \label{fig:feature_imp}
\end{figure}

First we applied XGB on the training set and we set the final value of the number of estimators to 12 where the loss function was the smallest measured on the validation set. We plot the predicted redshifts as the function of the spectroscopic redshifts as well as the residuals in Figure \ref{fig:photoz_XGB}. We can observe that the median prediction is very close to the spectroscopic value but the scatter is relatively large and there remained some bias in the residuals as well. At smaller redshifts this bias is mostly positive, which means that many of closeby ($z < 1$) quasars have been considered by XGB as distant objects. We also plotted the feature importance in Figure \ref{fig:feature_imp}. These values are calculated based on the so called information gain, meaning the average training loss reduction gained when using a feature for splitting. According to the results it seems that the near-infrared range is the most informative for XGB. To understand the high relevance of the \texttt{y\_w1\_dered} feature we calculated the median value for each of the standardized features along the redshift. We then plot the used redshift bin centers as a function of the median values of features (see Figure \ref{fig:colind_z}). We can observe that many of the features have large fluctuations which means that the same feature values (color indices) relate to several redshift ranges and therefore the correct prediction needs a lot of splits in these feature domains. Contrarily in case of \texttt{y\_w1\_dered} we can see that there is a broad redshift region where there is a one-to-one relation between the feature and the redshift. 

\subsection{Application of ANN}

\begin{figure}
	\includegraphics[width=\columnwidth]{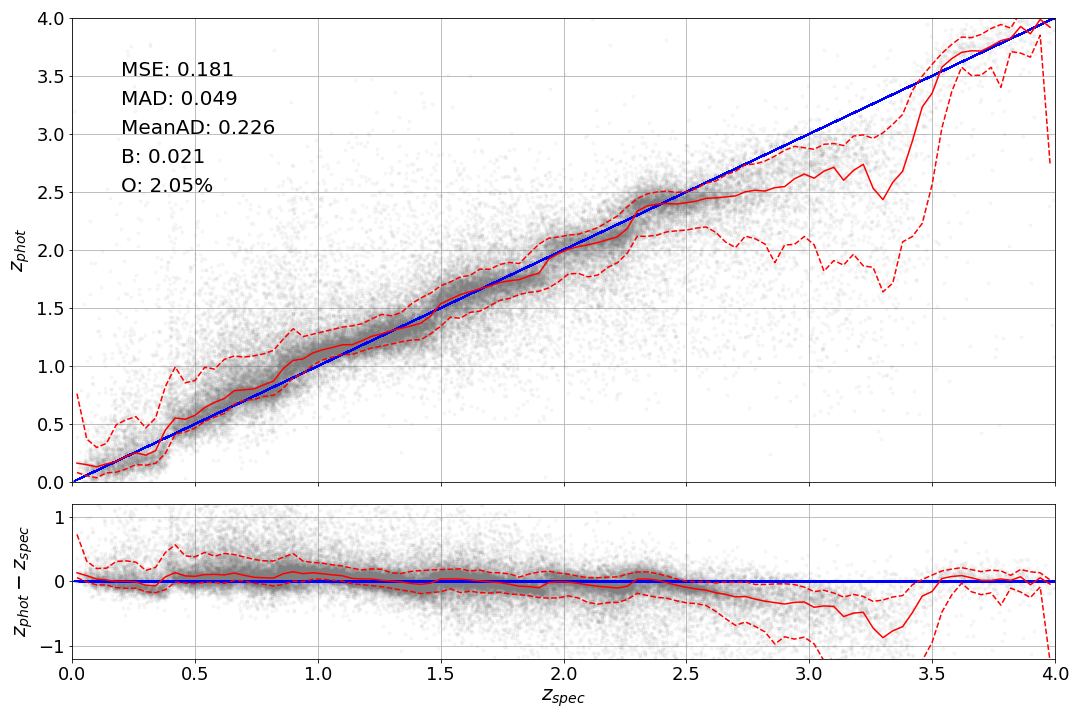}
    \caption{Photometric redshift estimates of ANN and the residuals. The red continuous and dashed lines refer to the median and the 68$\%$ confidence interval, respectively.}
    \label{fig:photoz_ANN}
\end{figure}

We then applied ANN on the data where we found better results after a few thousand iterations using a batch size of 1024. The results can be seen in Figure \ref{fig:photoz_ANN}. The plot is similar to Figure \ref{fig:photoz_XGB}, however the scatter is now narrower. We achieved a mean squared error of 0.18, and a mean absolute difference of 0.226 that is very similar to the result found in \cite{jin2019} (their MeanAD was 0.22). The characteristic fluctuation around the ground truth value remained there however, similarly to the results in \cite{jin2019}.

\subsection{Error estimation}

\begin{figure}
	\includegraphics[width=\columnwidth]{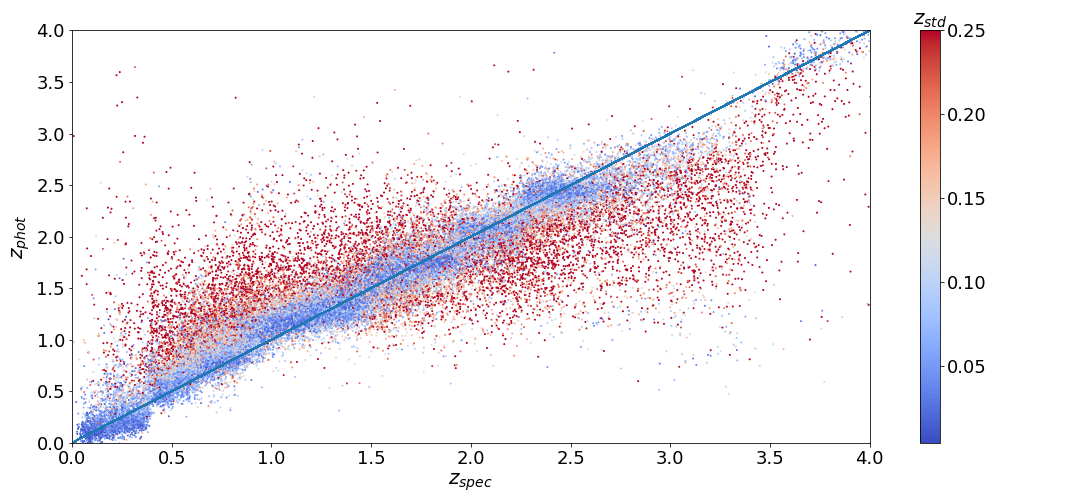}
    \caption{Photometric redshift predictions of ANN with error estimation.}
    \label{fig:photoz_ANN_with_err}
\end{figure}

To estimate the error of the ANN model we used a Monte Carlo approach. We randomly perturbed the magnitudes by adding a Gaussian distributed random variable with zero mean and standard deviation equal to the provided magnitude error. We then recalculated the color indices and applied the ANN on the perturbed data. We repeated this process 100 times and took the mean value as the final prediction and the standard deviation as the error of the estimated redshift. In Figure \ref{fig:photoz_ANN_with_err} we plotted again the results made on the test set but now the estimated error has been included as color coding. We can observe that the uncertainty is consistent with the accuracy of the predictions. Less uncertain photo-z estimations are closer to the spectroscopic redshift values.

\subsection{Explanation of the step-like structure}

\begin{figure}
	\includegraphics[width=\columnwidth]{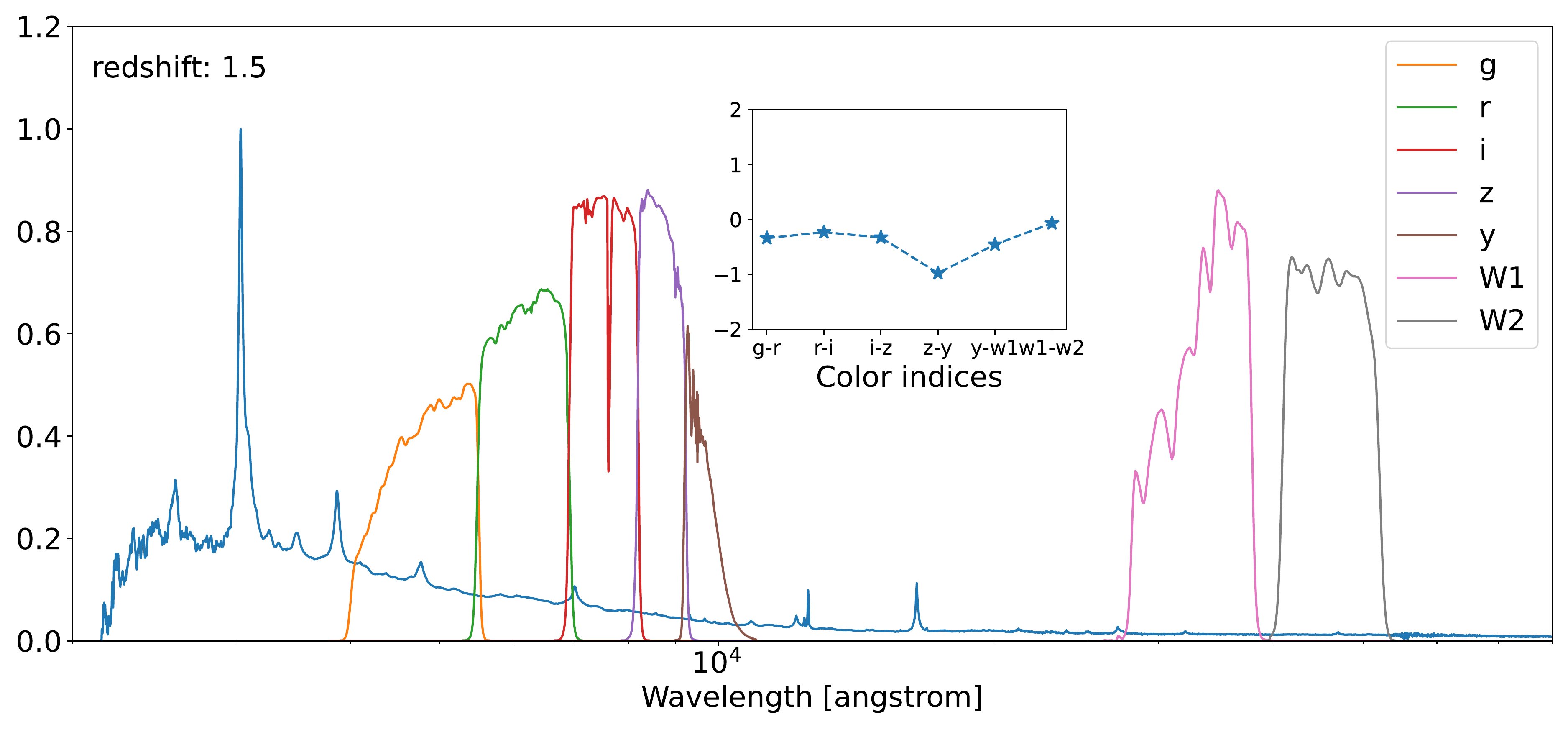}
    \caption{Composite quasar spectra at $z=1.5$, transmission curves of the PanSTARRS and WISE passbands as well as the determined synthetic color indices.}
    \label{fig:comp_spec}
\end{figure}

Now we give an explanation for this step-like structure of the diagram. First of all we need to recap the basic idea behind the usage of color indices in the prediction of the redshift. The color indices provide information about the flux ratio between the neighbouring photometric passbands. While the different emission lines of the quasar spectrum are moving to longer wavelengths during redshifting, the color indices will also change. However, since the quasars have less features in their continuum spectra compared to galaxies, the change in the color indices will be typically smaller than the level of photometric error, and therefore the model cannot catch the relation. The model performance will be better only if the color change is significant which occurs when one strong emission line goes from one passband to another. These relatively large changes occur only at specific redshifts and therefore the Machine Learning models will be first able to predict the corresponding redshift interval of the quasars. Now, since we try to minimize the mean squared error during the optimization process the model will predict in most cases the middle of the redshift interval that produces the smallest error for every quasar in that redshift interval. Hence, the resulting plot will contain several "steps" between the mentioned redshifts. To demonstrate our concept we used a composite quasar spectrum for the optical and near infrared regime downloaded from the website of the Space Telescope Science Institute (STScI)\footnote{https://www.stsci.edu/hst/instrumentation/reference-data-for-calibration-and-tools/astronomical-catalogs/composite-qso-spectra-for-nir}. In Figure \ref{fig:comp_spec} we plotted this spectrum, the transmission curves of the PS1 and WISE passbands and the calculated color indices.

Using these data we calculated the six color indices at 4000 redshift values in the range of $z \in [0,4]$. Then we calculated the standard deviation of each color index using a bin size of 0.04. Finally, we took the maximum standard deviation for each bin and plotted the results in Figure \ref{fig:photoz_ANN}. It can be clearly seen that the jumps in the redshift predictions occur at the same positions where the standard deviation in the color indices is relatively high which confirms our assumption.

\begin{figure}
	\includegraphics[width=\columnwidth]{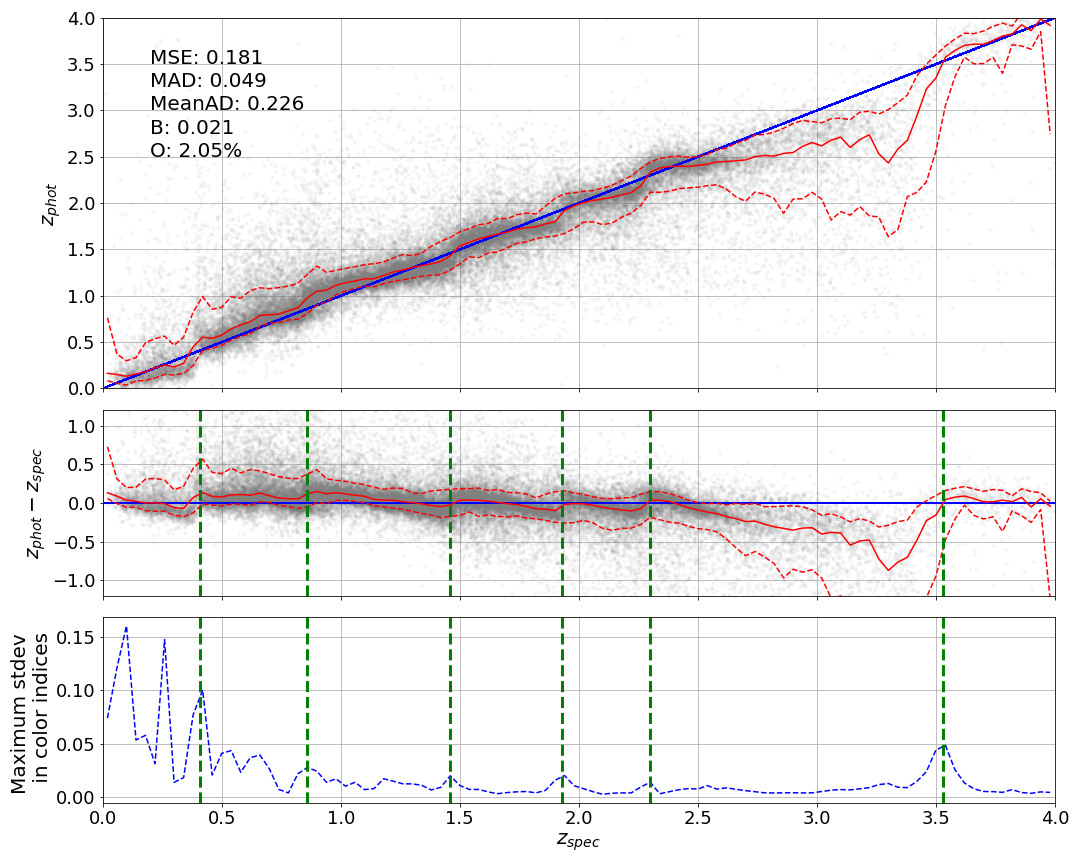}
    \caption{Photometric redshift estimation of ANN as well as the standard deviation of synthetic quasar color indices (blue dashed line) as a function of redshift. The positions of local maxima are marked with the green dashed vertical lines.}
    \label{fig:photoz_colind}
\end{figure}

\subsection{Independent validation of the results}

\begin{figure*}
\begin{center}
\includegraphics[draft=false,width=\textwidth]{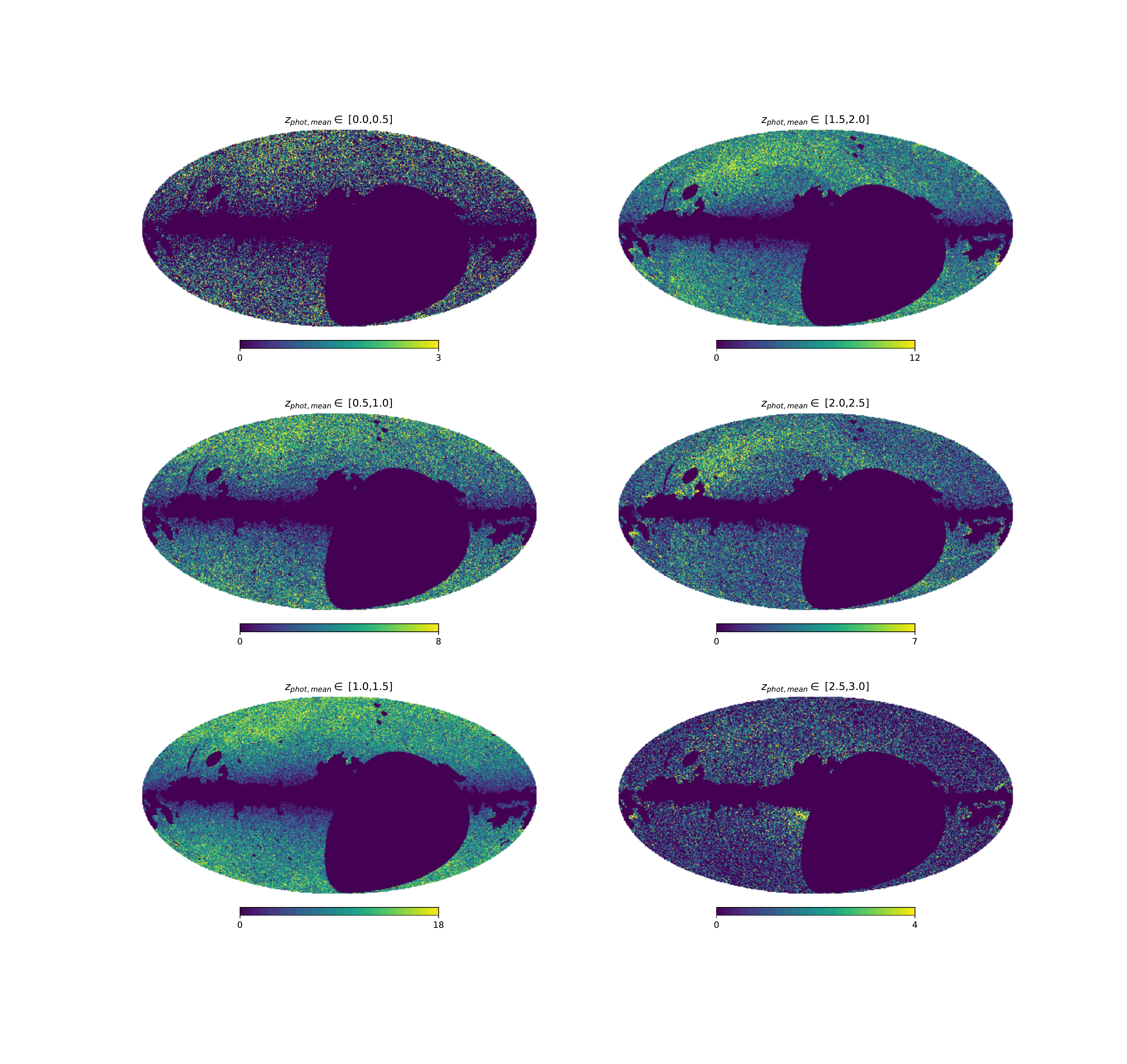}
\vspace*{-0.5cm}
\end{center}
\caption{Number count of quasars in the different redshift slices. The WMAP DR4 temperature analysis exclusion mask was used.}
\label{fig:hpx_maps}
\end{figure*}

\begin{figure*}
\begin{center}
\includegraphics[draft=false,width=\textwidth]{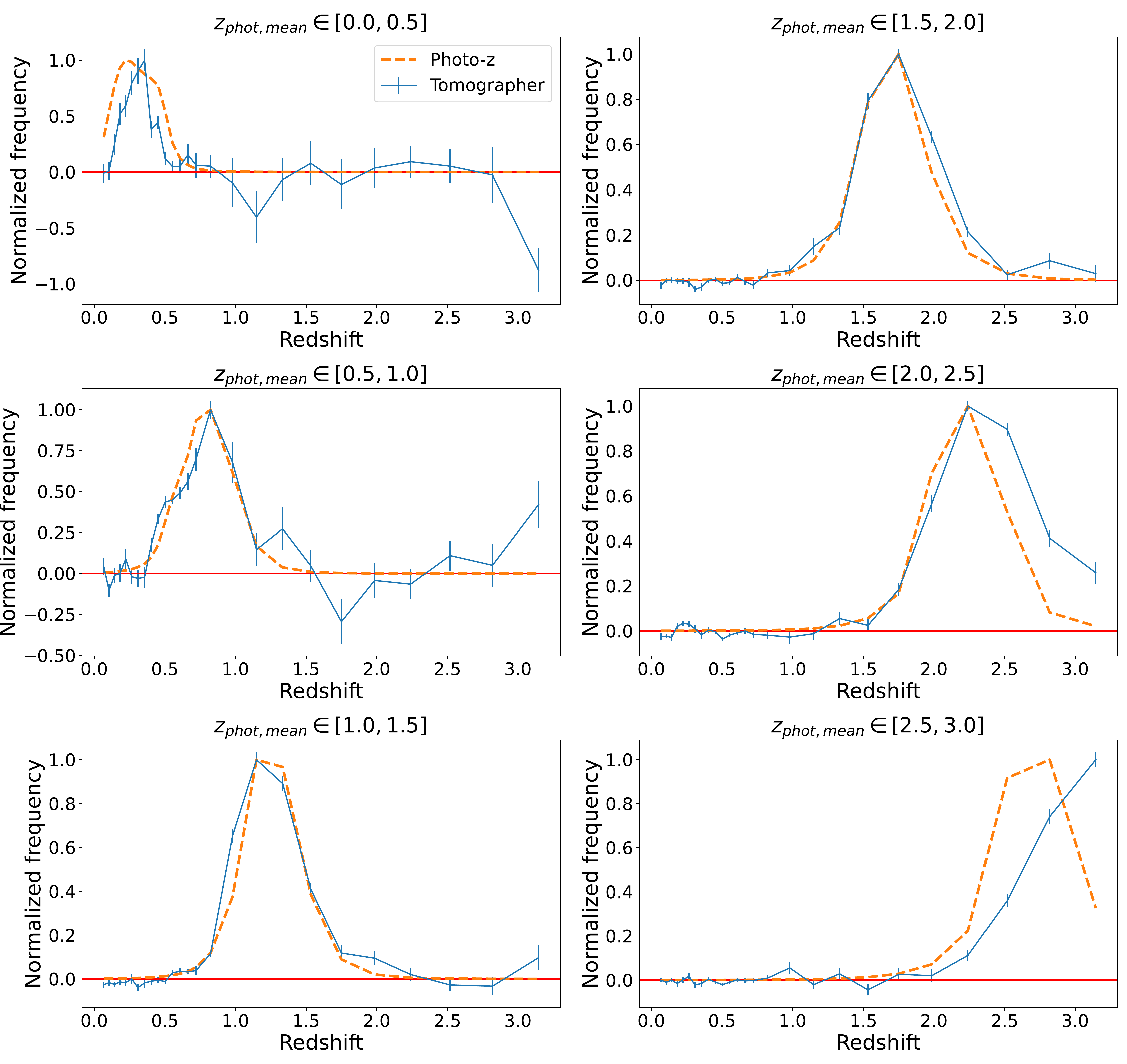}
\vspace*{-0.5cm}
\end{center}
\caption{Prediction of redshift dependence of frequency distribution calculated by Tomographer related to the different redshift slices (blue continuous line with errorbars). For comparison the frequency distribution of quasars along the photometric redshift has been plotted as well (orange dashed line).}
\label{fig:correlations}
\end{figure*}

In this section we demonstrate the reliability of the derived \textit{non-extrapolated} photometric redshifts by applying a completely independent algorithm, namely the so called clustering-based redshift estimation \cite{menard2013}. This approach uses a set of spatial cross-correlations between a photometric and a reference spectroscopic sample. This means that we need to provide a map of objects within a defined photometric redshift range and let the model to figure out the number density change as a function of redshift ($\frac{dN}{dz}$). We have also calculated the frequency distribution of the quasars with respect to the predicted photometric redshifts. We accounted for the estimated uncertainty of the predictions where we used a Gaussian distribution. We consider the outcome as a successful validation if the two distributions are close to each other. We used the publicly available Tomographer\footnote{tomographer.org} web user interface for this validation process. We created HEALPIX images about the spatial distributions of quasars in the galactic coordinate system using the \texttt{healpy} python package with NSIDE=128 and the WMAP DR4 temperature analysis exclusion mask \footnote{https://lambda.gsfc.nasa.gov/product/wmap/dr4/masks\_get.html}. We created these maps in 0.5 wide redshift bins and the resulting plots can be seen in Figure \ref{fig:hpx_maps}. The correlation results of Tomographer are plotted in Figure \ref{fig:correlations}. We can observe that Tomographer predicts such correlation values to its reference data set that are distributed very similarly to the frequency distribution of quasars calculated along the photometric redshifts, especially for $z < 2.5$. This confirms the reliability of the determined photometric redshift catalog. Only for the last $z\in[2.5,3.]$ redshift bin Tomographer predicts significantly larger redshift values, which is not surprising. Regarding to Figure \ref{fig:photoz_ANN} we can notice that just in that specific redshift range the model predicts systematically lower redshifts than the real values, and therefore a significant amount of distant quasars will fall into the $z\in[2.5,3.]$ photometric redshift bin.

\section{Conclusions}\label{conclusions}

We created a photometric redshift catalog for a total number of 4,849,611 quasars including error estimation. From these, 2,879,298 quasars are within the training set coverage and therefore the redshift estimations are more reliable for them. We presented an XGBoost machine learning model as a base line method and used a more advanced artificial neural network model for the final predictions. We provided a detailed analysis of the results and an explanation for the observed bias in the data. Finally, we validated our redshift catalog using a completely independent, clustering-based redshift estimation method. We found good accordance between the results of the two methods below $z < 2.5$ therefore the published catalog will be useful for several cosmological large-scale structure studies.

\section*{Acknowledgements}

This work was supported by the  Ministry of Innovation and Technology NRDI Office grants OTKA NN 129148 and the MILAB Artificial Intelligence National Laboratory Program. IS acknowledges support from the National Science Foundation (NSF) award 1616974.

%%%%%%%%%%%%%%%%%%%%%%%%%%%%%%%%%%%%%%%%%%%%%%%%%%
\section*{Data Availability}

The derived photometric redshift catalogue is publicly available at https://doi.org/10.5281/zenodo.6609756.



\bibliographystyle{mnras}
\bibliography{references} % if your bibtex file is called example.bib

\begin{thebibliography}{}
\makeatletter
\relax
\def\mn@urlcharsother{\let\do\@makeother \do\$\do\&\do\#\do\^\do\_\do\%\do\~}
\def\mn@doi{\begingroup\mn@urlcharsother \@ifnextchar [ {\mn@doi@}
  {\mn@doi@[]}}
\def\mn@doi@[#1]#2{\def\@tempa{#1}\ifx\@tempa\@empty \href
  {http://dx.doi.org/#2} {doi:#2}\else \href {http://dx.doi.org/#2} {#1}\fi
  \endgroup}
\def\mn@eprint#1#2{\mn@eprint@#1:#2::\@nil}
\def\mn@eprint@arXiv#1{\href {http://arxiv.org/abs/#1} {{\tt arXiv:#1}}}
\def\mn@eprint@dblp#1{\href {http://dblp.uni-trier.de/rec/bibtex/#1.xml}
  {dblp:#1}}
\def\mn@eprint@#1:#2:#3:#4\@nil{\def\@tempa {#1}\def\@tempb {#2}\def\@tempc
  {#3}\ifx \@tempc \@empty \let \@tempc \@tempb \let \@tempb \@tempa \fi \ifx
  \@tempb \@empty \def\@tempb {arXiv}\fi \@ifundefined
  {mn@eprint@\@tempb}{\@tempb:\@tempc}{\expandafter \expandafter \csname
  mn@eprint@\@tempb\endcsname \expandafter{\@tempc}}}

\bibitem[\protect\citeauthoryear{Abate et~al.,}{Abate et~al.}{2012}]{lsst}
Abate A.,  et~al., 2012, Research report, {Large Synoptic Survey Telescope:
  Dark Energy Science Collaboration}, \url
  {http://hal.in2p3.fr/in2p3-00748540}.
{IN2P3}, \url {http://hal.in2p3.fr/in2p3-00748540}

\bibitem[\protect\citeauthoryear{Beck, Dobos, Budavári, Szalay  \&
  Csabai}{Beck et~al.}{2016}]{beck1}
Beck R.,  Dobos L.,  Budavári T.,  Szalay A.~S.,   Csabai I.,  2016, \mn@doi
  [Monthly Notices of the Royal Astronomical Society] {10.1093/mnras/stw1009},
  460, 1371

\bibitem[\protect\citeauthoryear{Benitez}{Benitez}{2000}]{Benitez2000}
Benitez N.,  2000, \mn@doi [The Astrophysical Journal] {10.1086/308947}, 536,
  571

\bibitem[\protect\citeauthoryear{{Blanton} et~al.,}{{Blanton}
  et~al.}{2017}]{sdss1}
{Blanton} M.~R.,  et~al., 2017, \mn@doi [\aj] {10.3847/1538-3881/aa7567}, \href
  {https://ui.adsabs.harvard.edu/abs/2017AJ....154...28B} {154, 28}

\bibitem[\protect\citeauthoryear{{Bolzonella}, {Miralles}  \&
  {Pell{\'o}}}{{Bolzonella} et~al.}{2000}]{bolzonella}
{Bolzonella} M.,  {Miralles} J.~M.,   {Pell{\'o}} R.,  2000, \aap, \href
  {https://ui.adsabs.harvard.edu/abs/2000A&A...363..476B} {363, 476}

\bibitem[\protect\citeauthoryear{Boris, L.~Sodre, Cypriano, Santos, de Oliveira
   \& West}{Boris et~al.}{2007}]{Boris2007}
Boris N.~V.,  L.~Sodre J.,  Cypriano E.~S.,  Santos W.~A.,  de Oliveira C.~M.,
   West M.,  2007, \mn@doi [The Astrophysical Journal] {10.1086/519992}, 666,
  747

\bibitem[\protect\citeauthoryear{Brammer, van Dokkum  \& Coppi}{Brammer
  et~al.}{2008}]{Brammer2008}
Brammer G.~B.,  van Dokkum P.~G.,   Coppi P.,  2008, \mn@doi [The Astrophysical
  Journal] {10.1086/591786}, 686, 1503

\bibitem[\protect\citeauthoryear{Budav{\'{a}}ri}{Budav{\'{a}}ri}{2009}]{Budavari2009}
Budav{\'{a}}ri T.,  2009, \mn@doi [The Astrophysical Journal]
  {10.1088/0004-637x/695/1/747}, 695, 747

\bibitem[\protect\citeauthoryear{Carliles, Budav{\'{a}}ri, Heinis, Priebe  \&
  Szalay}{Carliles et~al.}{2010}]{Carliles2010}
Carliles S.,  Budav{\'{a}}ri T.,  Heinis S.,  Priebe C.,   Szalay A.~S.,  2010,
  \mn@doi [The Astrophysical Journal] {10.1088/0004-637x/712/1/511}, 712, 511

\bibitem[\protect\citeauthoryear{{Chambers} et~al.,}{{Chambers}
  et~al.}{2016}]{panstarrs}
{Chambers} K.~C.,  et~al., 2016, arXiv e-prints, \href
  {https://ui.adsabs.harvard.edu/abs/2016arXiv161205560C} {p. arXiv:1612.05560}

\bibitem[\protect\citeauthoryear{Chen \& Guestrin}{Chen \&
  Guestrin}{2016}]{xgb}
Chen T.,  Guestrin C.,  2016, in Proceedings of the 22nd ACM SIGKDD
  International Conference on Knowledge Discovery and Data Mining. KDD '16.
Association for Computing Machinery, New York, NY, USA, p. 785–794,
  \mn@doi{10.1145/2939672.2939785}, \url
  {https://doi.org/10.1145/2939672.2939785}

\bibitem[\protect\citeauthoryear{Coe, Ben{\'{\i}}tez, S{\'{a}}nchez, Jee,
  Bouwens  \& Ford}{Coe et~al.}{2006}]{Coe2006}
Coe D.,  Ben{\'{\i}}tez N.,  S{\'{a}}nchez S.~F.,  Jee M.,  Bouwens R.,   Ford
  H.,  2006, \mn@doi [The Astronomical Journal] {10.1086/505530}, 132, 926

\bibitem[\protect\citeauthoryear{Csabai, Connolly, Szalay  \&
  Budav{\'{a}}ri}{Csabai et~al.}{2000}]{Csabai2000}
Csabai I.,  Connolly A.~J.,  Szalay A.~S.,   Budav{\'{a}}ri T.,  2000, \mn@doi
  [The Astronomical Journal] {10.1086/301159}, 119, 69

\bibitem[\protect\citeauthoryear{Elliott, {de Souza}, Krone-Martins, Cameron,
  Ishida  \& Hilbe}{Elliott et~al.}{2015}]{elliott}
Elliott J.,  {de Souza} R.,  Krone-Martins A.,  Cameron E.,  Ishida E.,   Hilbe
  J.,  2015, \mn@doi [Astronomy and Computing]
  {https://doi.org/10.1016/j.ascom.2015.01.002}, 10, 61

\bibitem[\protect\citeauthoryear{Friedman}{Friedman}{2001}]{gbdt}
Friedman J.~H.,  2001, \mn@doi [The Annals of Statistics]
  {10.1214/aos/1013203451}, 29, 1189

\bibitem[\protect\citeauthoryear{{Harikane} et~al.,}{{Harikane}
  et~al.}{2022}]{distantgal}
{Harikane} Y.,  et~al., 2022, \mn@doi [\apj] {10.3847/1538-4357/ac53a9}, \href
  {https://ui.adsabs.harvard.edu/abs/2022ApJ...929....1H} {929, 1}

\bibitem[\protect\citeauthoryear{Hogan, Fairbairn  \& Seeburn}{Hogan
  et~al.}{2015}]{hogan}
Hogan R.,  Fairbairn M.,   Seeburn N.,  2015, \mn@doi [Monthly Notices of the
  Royal Astronomical Society] {10.1093/mnras/stv430}, 449, 2040

\bibitem[\protect\citeauthoryear{{Ilbert, O.} et~al.,}{{Ilbert, O.}
  et~al.}{2006}]{ilbert}
{Ilbert, O.} et~al., 2006, \mn@doi [A\&A] {10.1051/0004-6361:20065138}, 457,
  841

\bibitem[\protect\citeauthoryear{Jin, Zhang, Zhang, Zhao, Wu  \& Fan}{Jin
  et~al.}{2019}]{jin2019}
Jin X.,  Zhang Y.,  Zhang J.,  Zhao Y.,  Wu X.-b.,   Fan D.,  2019, \mn@doi
  [Monthly Notices of the Royal Astronomical Society] {10.1093/mnras/stz680},
  485, 4539

\bibitem[\protect\citeauthoryear{Krone-Martins, Ishida  \& de
  Souza}{Krone-Martins et~al.}{2014}]{krone}
Krone-Martins A.,  Ishida E. E.~O.,   de Souza R.~S.,  2014, \mn@doi [Monthly
  Notices of the Royal Astronomical Society: Letters] {10.1093/mnrasl/slu067},
  443, L34

\bibitem[\protect\citeauthoryear{Leistedt, Mortlock  \& Peiris}{Leistedt
  et~al.}{2016}]{leistadt}
Leistedt B.,  Mortlock D.~J.,   Peiris H.~V.,  2016, \mn@doi [Monthly Notices
  of the Royal Astronomical Society] {10.1093/mnras/stw1304}, 460, 4258

\bibitem[\protect\citeauthoryear{Liu, Moore  \& Gray}{Liu
  et~al.}{2006}]{balltree}
Liu T.,  Moore A.~W.,   Gray A.,  2006, J. Mach. Learn. Res., 7, 1135–1158

\bibitem[\protect\citeauthoryear{Lyke et~al.,}{Lyke et~al.}{2020}]{Lyke2020}
Lyke B.~W.,  et~al., 2020, \mn@doi [The Astrophysical Journal Supplement
  Series] {10.3847/1538-4365/aba623}, 250, 8

\bibitem[\protect\citeauthoryear{Magnier et~al.,}{Magnier
  et~al.}{2020a}]{Magnier2020}
Magnier E.~A.,  et~al., 2020a, \mn@doi [The Astrophysical Journal Supplement
  Series] {10.3847/1538-4365/abb829}, 251, 3

\bibitem[\protect\citeauthoryear{Magnier et~al.,}{Magnier
  et~al.}{2020b}]{Magnier2020c}
Magnier E.~A.,  et~al., 2020b, \mn@doi [The Astrophysical Journal Supplement
  Series] {10.3847/1538-4365/abb82c}, 251, 5

\bibitem[\protect\citeauthoryear{Magnier et~al.,}{Magnier
  et~al.}{2020c}]{Magnier2020b}
Magnier E.~A.,  et~al., 2020c, \mn@doi [The Astrophysical Journal Supplement
  Series] {10.3847/1538-4365/abb82a}, 251, 6

\bibitem[\protect\citeauthoryear{{M{\'e}nard}, {Scranton}, {Schmidt},
  {Morrison}, {Jeong}, {Budavari}  \& {Rahman}}{{M{\'e}nard}
  et~al.}{2013}]{menard2013}
{M{\'e}nard} B.,  {Scranton} R.,  {Schmidt} S.,  {Morrison} C.,  {Jeong} D.,
  {Budavari} T.,   {Rahman} M.,  2013, arXiv e-prints, \href
  {https://ui.adsabs.harvard.edu/abs/2013arXiv1303.4722M} {p. arXiv:1303.4722}

\bibitem[\protect\citeauthoryear{Miles, Freitas  \& Serjeant}{Miles
  et~al.}{2007}]{miles}
Miles N.,  Freitas A.,   Serjeant S.,  2007, in Ellis R.,  Allen T.,   Tuson
  A.,  eds, Applications and Innovations in Intelligent Systems XIV. Springer
  London, London, pp 75--87

\bibitem[\protect\citeauthoryear{{Nakoneczny, S. J.} et~al.,}{{Nakoneczny, S.
  J.} et~al.}{2021}]{kidscatalog}
{Nakoneczny, S. J.} et~al., 2021, \mn@doi [A\&A] {10.1051/0004-6361/202039684},
  649, A81

\bibitem[\protect\citeauthoryear{O'Mill, Duplancic, García~Lambas  \&
  Sodré}{O'Mill et~al.}{2011}]{omill}
O'Mill A.~L.,  Duplancic F.,  García~Lambas D.,   Sodré L.,  2011, \mn@doi
  [Monthly Notices of the Royal Astronomical Society]
  {10.1111/j.1365-2966.2011.18222.x}, 413, 1395

\bibitem[\protect\citeauthoryear{Schlafly et~al.,}{Schlafly
  et~al.}{2014}]{Schlafly2014}
Schlafly E.~F.,  et~al., 2014, \mn@doi [The Astrophysical Journal]
  {10.1088/0004-637x/789/1/15}, 789, 15

\bibitem[\protect\citeauthoryear{{The Dark Energy Survey Collaboration}}{{The
  Dark Energy Survey Collaboration}}{2005}]{DES}
{The Dark Energy Survey Collaboration} 2005, The Dark Energy Survey,
  \mn@doi{10.48550/ARXIV.ASTRO-PH/0510346}, \url
  {https://arxiv.org/abs/astro-ph/0510346}

\bibitem[\protect\citeauthoryear{{Tonry} et~al.,}{{Tonry} et~al.}{2012}]{tonry}
{Tonry} J.~L.,  et~al., 2012, \mn@doi [\apj] {10.1088/0004-637X/750/2/99},
  \href {https://ui.adsabs.harvard.edu/abs/2012ApJ...750...99T} {750, 99}

\bibitem[\protect\citeauthoryear{Wadadekar}{Wadadekar}{2005}]{wadadekar}
Wadadekar Y.,  2005, Publications of the Astronomical Society of the Pacific,
  117, 79

\bibitem[\protect\citeauthoryear{Waters et~al.,}{Waters
  et~al.}{2020}]{Waters2020}
Waters C.~Z.,  et~al., 2020, \mn@doi [The Astrophysical Journal Supplement
  Series] {10.3847/1538-4365/abb82b}, 251, 4

\bibitem[\protect\citeauthoryear{Wright et~al.,}{Wright
  et~al.}{2010}]{Wright2010}
Wright E.~L.,  et~al., 2010, \mn@doi [The Astronomical Journal]
  {10.1088/0004-6256/140/6/1868}, 140, 1868

\bibitem[\protect\citeauthoryear{Wu \& Jia}{Wu \& Jia}{2010}]{wu}
Wu X.-B.,  Jia Z.,  2010, \mn@doi [Monthly Notices of the Royal Astronomical
  Society] {10.1111/j.1365-2966.2010.16807.x}, 406, 1583

\bibitem[\protect\citeauthoryear{Yang et~al.,}{Yang et~al.}{2017}]{Yang}
Yang Q.,  et~al., 2017, \mn@doi [The Astronomical Journal]
  {10.3847/1538-3881/aa943c}, 154, 269

\bibitem[\protect\citeauthoryear{York et~al.,}{York et~al.}{2000}]{York2000}
York D.~G.,  et~al., 2000, \mn@doi [The Astronomical Journal] {10.1086/301513},
  120, 1579

\makeatother
\end{thebibliography}

%\begin{thebibliography}{99}
%\end{thebibliography}

%%%%%%%%%%%%%%%%%%%%%%%%%%%%%%%%%%%%%%%%%%%%%%%%%%

%%%%%%%%%%%%%%%%% APPENDICES %%%%%%%%%%%%%%%%%%%%%

%%%%%%%%%%%%%%%%%%%%%%%%%%%%%%%%%%%%%%%%%%%%%%%%%%

% Don't change these lines
\bsp	% typesetting comment
\label{lastpage}
\end{document}